\documentclass[12pt]{article}

\begin{document}                                                     


\def\g{{\tt g}}
\def\1{{\bf 1}}
\def\Z{{\bf Z}}
\def\ee{\end{equation}}
\def\be{\begin{equation}}
\def\l{\label}
\def\D{{\cal D}}
\def\U{{\cal U}}
\def\sin{{\rm sin}}
\def\cos{{\rm cos}}
\def\f{{\bf \Phi}}
\def\v{\varphi}
\def\O{\bf {\cal O}}
\def\C{\bf C}
\def\C{\bf C}
\def\Q{{\cal Q}}
\def\G{{\cal G}}
\def\CP{\bf CP}
\def\e{\rm e}
\def\0{\nonumber}
\def\eea{\end{eqnarray}}
\def\bea{\begin{eqnarray}}
\def\Tr{\rm Tr}
\def\IR{\bf R}
\def\ZZ{\bf Z}
\def\T{\tau}
\def\ep{\epsilon}




\def\title#1{\centerline{\huge{#1}}}
\def\author#1{\centerline{\large{#1}}}
\def\address#1{\centerline{\it #1}}
\def\ack{{\bf Acknowledgments}$\quad$}
\def\Bibliography{\begin{thebibliography}}
\def\endbib{\end{thebibliography}}


\begin{titlepage}

\hfill ULB-TH/03-19


\title{On the Tensionless Limit of Bosonic Strings,}
\title{Infinite Symmetries and Higher Spins} 
\vspace{1.5cm}
\author{Giulio Bonelli}
\vspace{.3cm}
\address{
Physique Theorique et Mathematique - Universite Libre de Bruxelles}
\address{
Campus Plaine C.P. 231; B 1050 Bruxelles, Belgium}
\address{e-mail address: gbonelli@ulb.ac.be}

\begin{abstract}
In the tensionless limit of string theory on flat background all the massive tower
of states gets squeezed to a common zero mass level and the free theory is described by
an infinite amount of massless free fields with arbitrary integer high spin. 
We notice that in this situation the very notion of critical dimension gets lost,
the apparency of infinite global symmetries takes place, and
the closed tensionless string can be realized as a constrained subsystem 
of the open one in a natural way.
Moreover, we study the tensionless limit of the Witten's cubic sting field theory and
find that the theory in such a limit can be represented as an infinite set of free arbitrary higher
spin excitations plus an interacting sector involving their zero-modes only.
\end{abstract}

\tableofcontents



\end{titlepage}

\section{Introduction}

As far as our knowledge in Quantum Field Theory is concerned, 
field theories of massless particles enjoy
the richest symmetry structures, gauge symmetries.
Moreover, massless particle theories exhibit 
much nicer quantum properties than massive ones.

It is therefore interesting to study seriously the analogous
for String Theories in order to understand if richer symmetry structures
appear in the limit in which the string tension vanishes.
A first encouraging and inspiring result in such a direction is contained in 
the paper by D.Gross \cite{Gross}.

The definition of the tensionless limit that we take as our starting point
is the one in which all the massive tower of string states 
gets squeezed to a common zero point energy.
This corresponds to the limit $\alpha'\to\infty$ keeping fixed
the typical oscillators variables describing the string
harmonics structure.
This already presents the tensionless theory within
a much wider and symmetric picture where 
infinite global symmetries could
appear mixing all the oscillator modes.
Their actual realization is one of the results obtained in this letter.

Let us notice that the approach that we follow is different and most probably
inequivalent to the one 
initiated by A. Schild about null strings in \cite{S} where the tensionless
limit was considered at fixed $\sigma$-model field coordinates rather than at
fixed oscillator variables.

A much inspiring approach to the problem comes from section 1.5 of \cite{HT1},
where a possible link between a general approach to free higher spin theories and
the tensionless limit of free open strings was sketched.
In this paper we will make that picture more precise and we will extend it to
the interacting case.
It will turn out that the resulting formalism naturally fits with the tensionless limit
of String Field Theory suggesting a general scheme to extend the model
of \cite{HT1} to the interacting level.

One of the results that we will obtain is that in the tensionless regime the
stringy notion of critical dimension gets lost, since the nihilpotency of the BRST
charge is established for flat backgrounds of any dimension.
This means that one can consistently formulate tensionless string theories 
in any dimension, their (usual) tensile deformation being consistent 
at the critical value of space-time dimension only.
As a countercheck, notice that
in the tensionless limit, in fact, the manifest $SO({\rm dim}-2)$ multiplet structure of the
light-cone states are no more forced to add up and combine into representations 
of the massive little group $SO({\rm dim}-1)$ for all the higher massive
levels in the string tower.

Within the approach that we develop for the tensionless limit of string
theory, we will be able to identify the $\U^*(\g_O)$ invariance algebra 
of the free theory which is the 
real anti-hermitian restriction of the
universal covering of the Lie algebra 
$\g_O=w(1,{\rm dim}-1)\oplus sp(\infty)$, where $w(1,{\rm dim}-1)$ is 
the Lie algebra of the Weyl group generated 
by $iso(1,{\rm dim}-1)$ and spacetime 
dilatation\footnote{It is an open issue if this could be extended to the full
conformal algebra or not. See section 2.2.}. The elements in $\g_O$
are given bilinears in the string creator/annihilator operators. These transformations
act mixing and rearranging the level and the spin structure of the massless higher spin 
excitations of the theory. 

Moreover, we realize the tensionless limit of the free closed
bosonic string as a constrained sub-system of the open one, the constrain
being the level matching condition. We work out the form of the gauge
symmetries of the theory and single out an
infinite amount of global symmetries.

Finally, we calculate the
tensionless limit of the three-string vertex of the Cubic Open String
Field Theory and find (not surprisingly), that, once the string field is
expanded in ordinary field components, 
the interaction takes place among the infinite set of field zero-modes only.
This agrees with general expectations from higher spin field theory on flat spacetime.

\section{Free tensionless strings}

Let us devote this first section to a close plan discussion 
of the tensionless limit of free bosonic strings.
For later convenience we start from the free open strings
with free boundary conditions in Minkowski flat space.

\subsection{Open Tensionless Free Strings}

Notice that some of the following starting arguments concerning the open strings are already 
{\it in nuce} contained in \cite{HT1} which we take the freedom to 
expand and complete for sake of clarity
\footnote{Some related ideas already predate the Henneaux-Teitelboim approach,
as for example \cite{OS}, but without any clear connection with string theory
and a correct BRST analysis of the spectrum.}.
Moreover, the relation between the tensionless limit of string theory 
and higher spin theories has already been noticed in \cite{FS}.

The content of such a system is given by the string center of mass
variables $(x_\mu,p^\mu)$ and the infinite set of oscillators 
$(a_{n\mu},a^*_{n\mu})$, with $n>0$.
They satisfy the usual canonical commutation relations (CCRs)
\be
[x_\mu,p^\nu]=i\delta_\mu^\nu
\quad , \quad
[a_{n\mu},a^*_{m\nu}]=\eta_{\mu\nu}\delta_{nm}
\l{ccr}\ee
and the other commutators are vanishing.
The string coordinate and its conjugate momentum are ($\sigma\in [0,\pi]$)
$$
X_\mu(\sigma)=x_\mu +\sum_{n>0} \sqrt{\frac{2\alpha'}{n}}\left(a_{n\mu}+a_{n\mu}^*\right)cos(n\sigma),
$$ 
$$
P^\mu(\sigma)=\frac{p^\mu}{\pi} +
\frac{1}{i\pi}\sum_{n>0} \sqrt{\frac{n}{2\alpha'}}\left(a_{n\mu}-a_{n\mu}^*\right)cos(n\sigma).
$$

In terms of the oscillator variables the Virasoro constraints take the form
\be
L_n=-i\sqrt{2n\alpha'}p\cdot a_n
+
\sum_{m>0}\sqrt{m(m+n)}a^*_m\cdot a_{n+m}
-\frac{1}{2}\sum_{m=1}^{n-1}\sqrt{m(n-m)}
a_m\cdot a_{n-m}
\ee
for $n>0$,
\be
L_0=\alpha' p\cdot p + \sum_{n>0}na^*\cdot a_n
\quad {\rm and} \quad
L_{-n}=L^*_n
\ee
(where the dot indicates space-time scalar product).

The tensionless limit appears straightforwardly.
In fact, just by rescaling $L_0\to L_0/\alpha'$
and $L_n\to iL_n/\sqrt{2|n|\alpha'}$ (where $n\not=0$),
the Virasoro algebra in the $\alpha'\to\infty$ limit 
gets contracted to the much simpler algebra
\be
[L_0,L_n]=0
\quad, \quad
[L_n,L_m]=\delta_{n+m}L_0
\l{contracted}\ee
despite any previous central extension which scales away.
This implies that for the tensionless limit there is no notion of 
critical dimension (as a requirement that the central extension
of the constrain algebra should vanish)
\footnote{Actually, this should be checked for the full algebra including the ghost 
contributions, but the core result is all here. See in the following for the full treatment.}.
Notice that the algebra (\ref{contracted}) is satisfied by the leading 
orders in $\alpha'$ of the Virasoro generators, that is
$L_0=p\cdot p$, $L_n=p\cdot a_n$ and $L_{-n}=p\cdot a^*_n$.

Introducing the relative anticommuting ghosts 
\footnote{There is an armless $\sqrt{n}$ rescaling factor with respect to the 
usual ghost fields which are used in the tensile string theory, due to the 
rescaling of the constrains. Moreover there is some $\alpha'$ rescalings whose
precise form can be obtained by comparison with the ordinary tensile string BRST charge.}
$c_n$, $c_n^*$ and $c_0$ 
as well as the anti-ghosts $b$'s (normalized by $[c_0,b_0]_+=1$,
$[c_m,b^*_n]_+=\delta_{mn}$ and $[c^*_n,b_m]_+=\delta_{nm}$),
we can write the BRST operator which implement the above constrain algebra, namely
\be
\Q=\frac{1}{2}c_0L_0+
\sum_{n>0}\left[
c_nL_n^*+c_n^*L_n-2c^*_nc_nb_0\right] \quad \Q^2=0
\l{brst-gen}\ee
which squares to zero if (\ref{contracted}) is satisfied.

For the tensionless open string we have
\be
\Q_O=\frac{1}{2}c_0p^2+
\sum_{n>0}\left[
c_np\cdot a_n^*+c_n^*p\cdot a_n-2c^*_nc_nb_0\right]
\l{brst-open}\ee
This satisfies 
$$
\Q_O^2=0
$$
for any space-time dimension (therefore proving the absence of a critical dimension 
for the tensionless open string theory).
Hence it follows that the conformal anomaly, whose vanishing typically 
determines the critical dimension, is a specific property of the tensile strings.
Indeed, one can deform the tensionless constrains with quadratic 
terms in the oscillators and recover the tensile string with its
proper anomaly cocycle.
It is an interesting open point to check whether or not such a quadratic 
deformation is unique (once the closure of the constrain algebra is enforced)
and to study if other deformations are meaningful.

The relevant ghost number operator is
$$
\G_O=\frac{1}{2}(c_0b_0-b_0c_0)+
\sum_{n>0}\left[c_n^*b_n-b_n^*c_n\right],
$$
which is anti-hermitian (in contrast with $\Q$ which is hermitian).
With this convention, the Fock vacuum has ghost number $-1/2$.

Corresponding to the above structure we can define a free string field theory
where the states are built over the Fock vacuum $|0>$ annihilated 
by the oscillators $a_n^\mu$ (as well as the corresponding ghosts $c_n$,
$b_n$ and $b_0$). The free action is 
\be
S_O=\frac{1}{2}<\psi_O|\Q_O|\psi_O>, \quad {\rm where}\quad \G_O|\psi_O>=-\frac{1}{2}|\psi_O>
\l{openfreeaction}\ee
and can be expanded in the field components.
Extending the scheme developed in \cite{HT1} for finite sets 
of oscillators (and adapting the reality condition of the string field
to our case), the action (\ref{openfreeaction}) can be shown to 
consist of a superposition of free massless multi-tensors whose
indices symmetry structure is determined by the expansion of the 
string field in string creation oscillators.

\subsection{Symmetries of the Open Tensionless Free Strings}

As it is well known, the action (\ref{openfreeaction}) admits the gauge symmetries 
$
\delta|\Psi_O>=\Q_O|\Lambda>
$,
where $|\Lambda>$ has ghost number $-3/2$.
Moreover, there are also global symmetries given by the real unitary 
operators preserving the ghost number and the BRST charge $\Q_O$.
These have to be unitary to preserve the measure in the string field space
and real to preserve the reality of the string field.
Therefore, the related generators are 
the zero ghost number real
anti-hermitian operators commuting with $\Q_O$.
The action on the string field is
$
\delta|\Psi_O>=\Gamma |\Psi_O>
$
where $\Gamma=\Gamma^*=-\Gamma^\dagger$, $[\G_O, \Gamma]=0$ and $[\Q_O, \Gamma]=0$

These operators are generically characterized by an expansion of the form
\be
\Gamma=\sum_{l\geq 0} \Gamma_{(l)}^{[m_1\dots m_l][n_1\dots n_l]}
c_{m_1}\dots c_{m_l}b_{n_1}\dots b_{n_l}
\l{gexp}\ee
where $\Gamma_{(0)}$ is the pure oscillator part.
The problem of classifying the global symmetries of the tensionless free open string
amounts to solve in general for $\Gamma$ of the form (\ref{gexp}) such that
$[\Q_0,\Gamma]=0$ and then to enforce the real anti-hermiticity
constrains $\Gamma^*=\Gamma=-\Gamma^*$.
Here we will single out an infinite set of solutions of the above conditions and
will not pretend to solve in full generality such a problem.

We proceed by studying first the problem of finding the $\Gamma$s such that
the expansion (\ref{gexp}) stops at $l=1$. These will form naturally a Lie
algebra in the product form $\g$=[Space-Time]$\oplus$[Internal].
Once $\g$ will be determined, we can consider the universal covering algebra
$\U(\g)$ of $\g$ formed by all the words in the generators of $\g$ (modulo the 
commutation relations in $\g$). The elements of $\U(\g)$ still commute with 
$\Q_O$ and admit an expansion of the form (\ref{gexp}).
Therefore the real anti-hermitian elements in $\U(\g)$ will fit the definition
of generators of global symmetries. We will denote by $\U^*(\g)$ the resulting restriction.

Therefore, we now calculate the Lie algebra $\g$ for the tensionless free open
bosonic string. In order to do it, let us notice that the requirement that
the expansion (\ref{gexp}) stops at $l=1$ means that the commutator of
$\Gamma_{(0)}$ with the constrains is given by a linear combination 
with numerical coefficients of the constrains themselves.
Since the constrains are at most linear in the oscillators, then
$\Gamma_{(0)}$ has to be (at most) quadratic in the oscillators.

As far as the space-time symmetries are concerned, notice that 
the Poincare' generators are left as they are by the tensionless limit, since
they do not depend on $\alpha'$ from the very beginning,
i.e. we have
\be P_\mu=p_\mu \quad {\rm and}\quad
M_{\mu\nu}=\frac{1}{2}(x_\mu p_\nu -x_\nu p_\mu)-\frac{i}{2}\sum_{n>0}
(a^*_{n\mu}a_{n\nu}-a^*_{n\nu}a_{n\mu})
\l{lorgen}\ee
(such that $[M_{\mu\nu},\Q_O]=0$).
Moreover there is an additional spacetime dilatation generator
\footnote{This choice of the dilatation generator implements 
the vanishing scaling dimension of the string oscillator variables
as well as the proper scaling dimensions of the ghosts.}
$$ D=\frac{1}{2}\left[x\cdot p+p\cdot x\right]-2ic_0b_0+i\sum_{m>0} \left(b^*_mc_m-c^*_mb_m\right),
\quad [\Q_O,D]=0 $$ which completes a space-time Weyl group.

The reader might doubt whether
the complete conformal group could be the full spacetime symmetry
of the model or not.
This could be verified or by building the generators of the conformal boosts
or by implementing the spacetime conformal inversion as a symmetry of the BRST charge $\Q_O$.
We let this point as an open issue.

As far as the internal symmetry part is concerned, we find that
the constrain Lie algebra is acted on by an $sp(\infty)$ algebra.
It is easy to see that
the BRST charge $\Q_O$ 
commutes with the bilinears in the oscillators
\be
l_{(mn)}=a_m\cdot a_n -c_mb_n-c_nb_m
\quad
l^*_{(mn)}=a^*_m\cdot a^*_n +c^*_mb^*_n+c^*_nb^*_m
\l{rota}\ee
$$
h_{mn}=a^*_m\cdot a_n+c^*_mb_n+b^*_mc_n+\frac{{\rm dim}-2}{2}\delta_{mn}
$$
(where "dim" is the space-time dimension).

The above generators close to form the following algebra
\be
[l_{(mn)}, l_{(pq)}]=0 \quad
[l_{(mn)}, h_{pq}]=\delta_{np}l_{(mq)}+\delta_{mp}l_{(nq)}
\l{symal}\ee
$$
[h_{mn}, h_{pq}]=\delta_{np}h_{mq}-\delta_{mq}h_{pn}
$$
$$
[l^*_{(mn)}, l^*_{(pq)}]=0 \quad
[l^*_{(mn)}, h_{pq}]=-\delta_{nq}l^*_{(mp)}-\delta_{mq}l^*_{(np)}
$$
$$
[l_{(mn)},l^*_{(pq)}]= h_{qm}\delta_{np}+
h_{pm}\delta_{nq}+h_{qn}\delta_{mp}+h_{pn}\delta_{mq}
$$
which is $sp(\infty)$.

So we find that $\g_O=\left[w(1,{\rm dim}-1)\right]
\oplus\left[sp(\infty)\right]$ -- where $w(1,{\rm dim}-1)$ is the algebra
of the Weyl group generated by $iso(1,{\rm dim}-1)$ and the dilatation $D$ --
and we conclude that the elements in $\U^*(\g_O)$ are global 
symmetry generators.
It would be interesting to to study if the kind of algebras we are obtaining 
are related to the Borcherds ones \cite{B}.

An important
open point has to do with the full determination of the free theory
global symmetry algebra.
It would be interesting to check if the set of elements that we singled out is
all the global symmetry algebra or just a subalgebra.
With respect to this issue, let us notice that 
an infinite subset of global symmetry generators are given by the BRST exact ones,
namely the ones such that $\Gamma_e=\left[\Q_O,\Gamma_{-1}\right]_+$, with 
$\Gamma_{-1}$ an arbitrary real anti-hermitian operator of ghost number $-1$.
These transformations have to be considered as trivial and therefore 
the problem of calculating the global symmetries reduces to 
the solution of a BRST cohomology at zero ghost number.
This is well defined since, if $[\Gamma,\Q_O]=0$, then 
$\Gamma\Gamma_e=\Gamma\left[\Q_O,\Gamma_{-1}\right]_+=\left[\Q_O,\Gamma\Gamma_{-1}\right]_+$
is exact too and therefore exact generators form a (two side) ideal.

\subsection{Closed Tensionless Free Strings}

Another property which appears in the tensionless limit is a clear 
embedding of the closed string Hilbert space as a BRST invariant 
subspace in the open string unrestricted one.
This can be realized as follows. Let us write the closed string coordinates and
momenta by collecting the left/right moving oscillators in a common
set split in odd and even, that is (now $\sigma\in[0,2\pi]$)
$$
X^\mu(\sigma)=x^\mu+\sum_{n>0}\sqrt{\frac{\alpha'}{2n}}
\left(
a_{2n}e^{-in\sigma} +a_{2n+1}e^{in\sigma} 
+
a_{2n}^*e^{in\sigma} +a_{2n+1}^*e^{-in\sigma}
\right)
$$
$$
P_\mu(\sigma)=\frac{p_\mu}{2\pi}+\frac{1}{2\pi}
\sum_{n>0}\sqrt{\frac{n}{2\alpha'}}
\left(
-ia_{2n}e^{-in\sigma} -ia_{2n+1}e^{in\sigma} 
+
ia_{2n}^*e^{in\sigma} +ia_{2n+1}^*e^{-in\sigma}
\right)
$$
where the above modes satisfy the same CCRs (\ref{ccr}) as above.
Calculating the Virasoro constrains and performing a scaling similar
to the one that we already performed in the open string case, we find that 
the left over constrains for the tensionless closed string are 
$$ p^2=0, \quad p\cdot a_n=0, \quad  p\cdot a_n^*=0 \quad {\rm and}
\quad \sum_{n>0}(-1)^n n a^*_n\cdot a_n=0,$$
the last one being the level matching condition.

The BRST charge implementing the contracted closed string Virasoro algebra
is (we add here a further ghost/anti-ghost pair $c'_0$ and $b'_0$ for the 
level matching condition with $[c'_0,b'_0]_+=1$)
\be
\Q_C=\frac{1}{2}c_0p^2+
\sum_{n>0}\left(c_np\cdot a_n^*+c_n^*p\cdot a_n-2c^*_nc_nb_0\right)
+c'_0\sum_{n>0}n(-1)^n(a^*_na_n+b^*_nc_n+c^*_nb_n)
\l{brst-closed}\ee
which satisfies 
$$
\Q_C^2=0
$$
in any dimension. Notice that, with respect to eq.(\ref{brst-open}),
identifying the set of canonical coordinates with the open string one, we can rewrite
$$
\Q_C=\Q_O+c'_0 L
\quad {\rm where} \quad
L=\sum_{n>0}n(-1)^n(a^*_na_n+b^*_nc_n+c^*_nb_n)
$$
is the (ghost extended) level matching constrain which satisfies 
$$
[\Q_O,L]_-=0
$$
Therefore, one can interpret the closed tensionless string Hilbert space as 
an invariant subspace, singled out by the condition $L=0$ on the physical 
states, of the open tensionless string one.

Let us notice that we can write an action for the free tensionless 
closed string field theory, that is
$$
S_C=\frac{1}{2}<\psi_C|\Q_C|\psi_C>
$$
which can be written as a constrained open tensionless one.
By expanding $|\psi_C>=|\psi_O>+ c'_O|\phi_O>$ and substituting
$\Q_C=\Q_O+c'_0 L$, we calculate the Berezin integral over $c'_0$ in the scalar
product and find 
$$
S_C=\frac{1}{2}<\psi_O|L|\psi_O>-<\phi_O|\Q_O|\psi_O>
$$
The equations of motion corresponding to $S_C$ are in fact
$$
\Q_O|\psi_O>=0
\quad {\rm and} \quad
L|\psi_O>=\Q_O|\phi_O>
$$
which correctly implement the constrain $L=0$ in the BRST quantization scheme.

Let us point out that in fact such a construction embeds the free closed tensionless
string in the big Hilbert space of the open one.
Once the ghost number constrain $\G_C|\psi_C>=-\frac{1}{2}|\psi_C>$
is imposed, 
where in our notation $\G_C=\frac{1}{2}\left(c'_0b'_0-b'_0c'_0\right)+\G_O$,
then this becomes on the components
$$
\G_O|\psi_O>=0 \quad {\rm and}\quad \G_O|\phi_O>=-|\phi_O>
$$
which projects orthogonally to the $\G_O|\psi_O>=-\frac{1}{2}|\psi_O>$ condition.

The theory still admits gauge and global symmetries.
The gauge symmetries act as 
$$
\delta|\psi_O>=\Q_O|\Lambda_O>
\quad{\rm and}\quad
\delta|\phi_O>=L|\Lambda_O>-\Q_O|\Lambda'_O>
$$
where $|\Lambda_O>$ and $|\Lambda'_O>$ have ghost number $-1$ and $-2$
respectively.
As far as the infinite global symmetry (\ref{symal}) is concerned, let us point out that
it is partly broken by the level matching constrain and therefore 
only the subalgebra commuting with the shifted level matching combination 
$\sum_{n>0}n(-1)^n h_{nn}$
is present in the free tensionless closed string.
This can be easily calculated to be generated by the elements
\be
K_{mn}=l_{(2m+1,2n)}\quad
K^*_{mn}=l^*_{(2m+1,2n)}\quad
I_{mn}=h_{2m+1,2n+1} \quad
J_{mn}=h_{2m,2n}
\l{closymal}\ee
whose commutation relations can be worked out directly from (\ref{symal})
by simply substituting (\ref{closymal}).

\section{Interacting tensionless strings}

As we can see, in the tensionless limit the string coordinate field $X^\mu(\sigma)$
gets undone since it blows up.
From the interacting theory point of view this implies
the uncontrolled string fragmentation which was observed in such a regime
in \cite{Gross2}.
In few words, the tensionless string is unable to stabilize dynamically.
The string evolves as a incoherent set of massless particles
where the string profile is just constrained to be orthogonal 
to the c.m. momentum.
Such a picture is intrinsically unstable under interaction, as the results 
in \cite{Gross2} demonstrate, and the string profile gets undone.

It is therefore compelling to turn to an alternative picture to represent the
string interaction in the tensionless limit. 
The natural alternative picture is the string field interacting theory \cite{Witten}. 
In the following we study the tensionless limit of Witten's
cubic string field theory.

\subsection{The tensionless limit of cubic string field theory}

The interacting action at generic value of $\alpha'$ can be written as
\be
S_O=\frac{1}{2}<\Psi_O|\Q|\Psi_O>+\frac{g_o}{3}<V_3|_W\left(|\Psi_O>\otimes|\Psi_O>\otimes|\Psi_O>\right)
\l{csft}\ee
where $\Q$ is the full BRST charge and $<V_3|$ the three strings vertex 
\be
|V_3>_W={\cal N}\int dp^{(1)}dp^{(2)}dp^{(3)}
\delta\left(p^{(1)}+p^{(2)}+p^{(3)}\right)
\l{3sv}\ee
$$
{\rm exp}\left(
-\frac{1}{2}\sum_{1\leq r,s\leq 3}\left[
\sum_{m,n>0}V^{r,s}_{mn}a^{(r)*}_m\cdot a^{(s)*}_n+2\sqrt{\alpha'}\sum_{m>0}V^{r,s}_{m0}
a^{(r)*}_m\cdot p^{(s)}+\alpha'V^{r,s}_{00}p^{(r)}\cdot p^{(s)}
\right]\right)
$$ $$
\left(|p^{(1)}>\otimes|p^{(2)}>\otimes|p^{(3)}>\right)
$$
where the Neumann coefficients $V^{r,s}_{mn}$ are computable numbers (see,
e.g.\cite{GJ}) and ${\cal N}$ a normalization constant.

To obtain a picture of the three string vertex in which the tensionless limit
can be taken it is useful to pass from the c.m. momentum to the c.m. 
position representation. This amounts to insert a c.m. completition 
${\bf 1}_{c.m.}=\otimes_{s=1,2,3}
|x^{(s)}>\int dx^{(s)}<x^{(s)}|$
and to perform the Fourier transform by integrating over the momenta.
This can be done by first Fourier transforming the $\delta$-function
imposing momentum conservation so that the resulting integrals are Gaussians.
The resulting expression reads
$$
|V_3>_W={\cal N}
{\rm exp}\left(
-\frac{1}{2}\sum_{1\leq r,s\leq 3}
\sum_{m,n>0}V^{r,s}_{mn}a^{(r)*}_m\cdot a^{(s)*}_n\right)
$$ $$
\int dx^{(1)}dx^{(2)}dx^{(3)}
|x^{(1)}>\otimes |x^{(2)}>\otimes |x^{(3)}>
\int dx\frac{1}{(2\pi)^D}
\left(\frac{2\pi}{\alpha'}\right)^{\frac{3D}{2}}
\frac{1}{\sqrt{{\rm det}V_{00}}}
$$ $$
{\rm exp}\left(
\frac{1}{2\alpha'}
\left(\sqrt{\alpha'}a^{(s)*}_nV_{n0}^{sr}-ix^{(r)}-ix\right)
\left(V_{00}^{-1}\right)^{rq}
\left(\sqrt{\alpha'}a^{(t)*}_mV_{0m}^{qt}-ix^{(q)}-ix\right)
\right)
$$

We can perform the zero tension limit to the above expression
which gives 
$$
|V_3>_{WO}= {\cal N}_O (\int dx) 
{\rm exp}\left(
-\frac{1}{2}\sum_{1\leq r,s\leq 3}
\sum_{m,n>0}\hat V^{r,s}_{mn}a^{(r)*}_m\cdot a^{(s)*}_n\right)
\otimes_{s=1,2,3}\int dp^{(s)}\delta(p^{(s)})|p^{(s)}>
$$
where
$\hat V^{r,s}_{mn}=V^{r,s}_{mn}-\sum_{1\leq q,t\leq 3}
V_{m0}^{r,q}\left(V_{00}^{-1}\right)^{q,t}V_{0n}^{t,s}
$ and we have included some factors in the normalization constant
as ${\cal N}={\cal N}_O \frac{\sqrt{{\rm det}V_{00}}
  (\alpha')^{3D/2}}{(2\pi)^D}$.

As we see, in the tensionless limit the above three string vertex reduces to the 
zero momentum sector the interacting part of the theory and causes the
$*$-product to project on the zero mode sector the string fields.
This implies that the $*$-product in this limit does not admit an identity.

Therefore, in the tensionless limit the propagating degrees of freedom
remain free and stay decoupled from the zero momentum sector
where all the interaction occurs within the oscillators. 
As far as the interacting internal oscillator degrees of freedom are
concerned, the tensionless limit therefore reduces to a
zero dimensional interacting model very much similar to the matrix models
which are conjectured to be at the basis of an M-theory description.

In formulas, we have therefore
\be
S=\frac{1}{2}<\Psi_O|\Q_O|\Psi_O>+\frac{g_o}{3}<V_3|_{WO} |\Psi_O>^{\otimes 3}
\l{lll}\ee
In order to specify in a better way how the tensionless limit of the vertex
looks, let us rephrase the above results in a more abstract context.
As it is clear from the results obtained above,
the three string vertex in the tensionless limit satisfies the tensionless
limit of the string overlap conditions, that is
$$
(p_\mu^{(r)}+p_\mu^{(r+1)})|V_3>_{WO}=0 \quad {\rm and}
$$
$$
\sum_{n>0}\frac{1}{\sqrt{n}}\left[\left(a^{(r)}_{n\mu}+a^{*(r)}_{n\mu}\right)
+(-1)^n\left(a^{(r+1)}_{n\mu}+a^{*(r+1)}_{n\mu}\right)\right]
\cos(n\sigma)|V_3>_{WO}=0
$$
(where $\sigma\in [0,\pi/2]$)
and is therefore determined, up to a normalization, by these overlap conditions,
BRST invariance, reality condition, associativity of the induced $*$-product and cyclic symmetry.
Notice that the very reason for the projection on the zero center of mass momentum states in
the interaction term is the separation of oscillators and momenta in the
overlap conditions for the three (and higher) string vertex.

As far as the ghost oscillators factor is concerned, because of the relative $\alpha'$
rescaling in the ghost sector, this undergoes a treatment similar to the one
of the matter sector.
Moreover, the $\sqrt{n}$ factors that we use to rescale the ghost modes
enter the definition of the ghost part of the vertex. In our notation the
ghost part of the three string vertex is then
\be
{\rm exp}\left\{-\sum_{mn>0}\sum_{1\leq r,s\leq 3}
c^{*(r)}_m \hat V^{r,s}_{mn} b^{*(s)}_m\right\}
\l{ghve}\ee
as far as the ghost oscillators are concerned
and the ghost zero-mode overall factor $\prod_t c_0^{(t)}$.
With the above ghost completition, the vertex satisfies 
the condition
$\sum_{1\leq t\leq 3}\Q_O^{(t)}|V_3>_{WO}=0$
and has the correct $\G_O$ ghost number assignment (namely $1/2$ for each
entry) and satisfies the proper tensionless ghost overlap conditions.

Let us comment that the projection on the momentum zero-modes of the
$*$-product in the tensionless limit can be obtained also from 
the expanded for of the string field action.
A rough idea about this can be inferred e.g. from the expansion of the 
tachyon contribution to the cubic term of the action (see \cite{KS}), that is
$\int \tilde\phi^3$, where 
$\tilde\phi={\rm e}^{-\alpha'\ln(4/3\sqrt{3})\partial^2}\phi$.
The $\alpha'\to\infty$ limit of the above is well defined if 
the non constant part of $\phi$ scales away
and therefore $\int \tilde\phi^3 \to \int \phi_0^3$
where $\phi_0$ is the $\phi$ zero-mode.
At the same time, the tachyon mass parameter lifts to zero.

The result we obtained is in agreement with the general results of higher spin
theories \cite{nci} stating that on flat backgrounds there should not exist a consistent 
interaction scheme for higher spin fields.

Notwithstanding the non-dynamical form of the interaction terms which we
calculated in the last section, it would nonetheless make sense to check 
how much of the infinite symmetry of the free theory is preserved by the
interaction term. It seems that this problem should be studied in the
proper oscillator $\kappa$-basis where the $*$-product structure 
gets diagonalized to a flat Moyal product along the lines suggested in \cite{chu}.
From that point of view it is likely that an infinite invariance group
of symmetries of the flat symplectic structure arising in the tensionless case
is preserved.

\vspace{.5cm}

In the spirit of the previous discussion, it is possible in principle
to extend the string field method for the interactions to the higher spin
models considered in \cite{HT1} (the generic case
with arbitrary bosonic and fermionic roots of the Hamiltonian)
and to strings on consistent curved backgrounds.

The specific form of the interacting theory action, i.e. 
with a proper single three field vertex, is indicated by strong symmetry
arguments that apply more in general.
From a minimal point of view, let's assume the action to be given by
\be
S=\frac{1}{2}<\Psi|\Q|\Psi>+\frac{g}{3}<V_3||\Psi>^{\otimes 3}
\l{gia}\ee
where $<V_3|$ is a general three field vertex (i.e. an element of the third
tensorial product dual Hilbert space) which we take cyclic symmetric
and compatible with a given choice $R$ of reality condition 
\footnote{In OSFT this is the usual $Ra_n^\mu R=(-1)^na_n^\mu$
-- since the conjugation ${}^*$ is the composition of the hermitean conjugation
and BPZ conjugation -- 
(and similarly on the other variables) and the string zeromodes invariant,
but in general it is a possible choice up to equivalences.}
on the field $|\Psi>$ (which makes $\Q$ hermitian).
It is clear that if the $*$-product determined by $<V_3|$ and $R$ is 
associative and if 
$\Q$ acts as a graded differential with respect to $*$, then
the action (\ref{gia}) is invariant under the 
gauge transformations 
\be
|\Psi>\quad\to\quad |\Psi>+|\delta\Psi> \quad {\rm where}\quad |\delta\Psi>=\Q|\Lambda>
-g\left(|\Psi*\Lambda>-|\Lambda*\Psi>\right)
\l{trasf}\ee
Conversely, suppose we ask for a symmetry principle which extends the free
theory gauge invariance $|\delta\Psi>=\Q|\Lambda>$ to an interacting one.
Then, assuming (\ref{trasf}), then it is a symmetry of (\ref{gia}) if 
the above conditions are satisfied.
Notice that
this is independent on the specific nature of the vertex $<V_3|$.
By this we mean that a possible overlap condition scheme 
which could be posed to single out a given vertex, has not to be understood 
as necessary conditions, but rather 
should instead fit in a spectral scheme classifying the possible 
{\it graded differential ring} structures allowed by the Hilbert space of fields.

As we see, the scheme which underlies the introduction of interactions in string field
theory along with a deformation of the free theory gauge invariance
naturally extends to the higher spin models studied in \cite{HT1} and,
once it is extended to Anti-deSitter spaces,
resembles very much the approach in \cite{V}.
It would be interesting to translate such an approach in the language
of \cite{HS} too.

\section{Conclusions and open questions}

\paragraph{Relations with superstring theories}

Most of the above considerations extend also to superstrings.
Actually, one of the motivating issues which started the 
program of tensionless strings is also the challenge of understanding the
M-theory prevision about Little String Theory \cite{strominger}.  
It seems that here we have some of the ingredients to formulate
a string field theory for closed tensionless superstrings with a $U(N)$
symmetry in six dimensions, which we would like then to compare with the
expected properties of Little String Theory, i.e. with the tensionless closed
string theory which should describe the microscopic world-brane dynamics the
M5-branes bound states. There is still a basic point to understand in this
direction, namely if and how the string coupling constant is fixed to some
"self-dual" value and by which actual mechanism. This issue points directly to
the problem of a correct formulation of the interacting tensionless 
closed (super)string field theory
which we believe should develop within the scheme proposed in the previous sections.

Another issue to understand which comes naturally out of the above results 
(once we presume a similar behavior to hold for type II A superstrings) is 
whether the picture we obtained for string theory in the tensionless regime
can be understood as a decompactification of the matrix picture of M-theory
of \cite{BFSS}.
Notice that in such a limit it is no more true that D-particles decouple
because of infinitely massive in the weak coupling regime, the limit being 
$l_s\to 0$ at $g_s$ fixed. This corresponds to the eleven dimensional 
(wrapped membrane) tensionless limit $l_p\to 0$ along with $R_{11}/l_p=g_s^{2/3}$ fixed.

In principle one can consider other scalings in parallel to the tensionless limit.
For example, one could consider OSFT with a constant background B-field and
then perform a double scaling limit. This could give other interesting
possibilities, although with broken Poincare' invariance and non-commutative
space-time geometry.

Another feature of the tensionless limit is the lift of the tachyon states to 
the common massless level. This implies the stability of these string field theories
and shows that such a property is proper to the tensile deformation.
It could be interesting to understand if some tachyonic/unstable string sector 
can condense in a tensionless string field theory regime.
With this respect, it would be interesting to calculate the tensionless limits 
of open strings with different Dirichlet boundary conditions.
Since it introduces a further length scale,
this could also be seen as a toy model to test the effects of space-time curvature
on the tensionless string.

\paragraph{More general issues}

It is of clear importance a target space interpretation of the
whole tensionless string theory. In particular, the possibility of
characterizing the tensionless string in curved backgrounds has to be 
studied as well as the relation between the microscopic symmetries of the 
theory and the properties of the target space-time.
All this should be encoded in the very structure of the BRST operator $\Q$ and
the three string vertex (or the $*$-product) in a clear geometric way.
This extension should find wide application in gauge/string dualities
once Anti-deSitter backgrounds are concerned \cite{holo}.

A further open point has to do with a proper geometric interpretation of the 
zero tension limit from the world-sheet geometry point of view.
Actually, the contraction of the Virasoro algebra to the tensionless 
algebra, which is possible in general and not only in the free field
realization,  awaits a geometric conceptual interpretation which should replace the 
two dimensional conformal symmetry in such a limit.
The tensionless contraction of the Virasoro algebra is actually 
well defined in general and should be at the heart of the tensionless limit of string
theory on any consistent background.
Notice that with respect to (\ref{brst-gen}), we have $\Q^2=0$ for any
realization of the tensionless string algebra (\ref{contracted}).

It is not clear if the infinite symmetries that we have found in the
tensionless limit might somehow survive in the tensile regime.
If this would be the case, which a priori seems not given, this could
be a consequence of extra properties of the Neumann coefficients
such as the ones obtained in \cite{loriano}.
Actually this might have a counterpart in the construction in \cite{GW}.

The point of checking the global symmetries of the cubic interaction term 
remains an open one and can not be avoided in a complete analysis of the theory.

\vspace{.3cm}

It seems to the author that a lot of other basic questions regarding the tensionless string theory
and the meaning of its huge symmetry remain to be issued and answered. 
We hope that this note has driven the reader's attention to some expected and
unexpected properties of tensionless strings.

\vspace{.5cm}
{\bf Note added:} Recently the interesting paper \cite{LZ} appeard and its approach to the 
tensionless limit of string theory  on flat spacetime 
is similar in principle to the one considered in \cite{HT1} and here.
Let us notice also that the approach that we follow is in principle different to the
"null strings" one
initiated by A. Schild in \cite{S} where the tensionless
limit was considered at fixed $\sigma$-model field coordinates rather than at
fixed oscillator variables.

\vspace{.5cm}

\ack
I would like to thank G. Barnich, M. Bertolini, L. Bonora, N. Boulanger, M. Henneaux and
A. Sagnotti for stimulating discussions and encouragement.
\noindent
This work is supported by the Marie Curie fellowship contract
HPMF-CT-2002-0185.
Work supported in part by the ``Actions de Recherche
Concert{\'e}es" of the ``Direction de la Recherche Scientifique -
Communaut{\'e} Fran{\c c}aise de Belgique", by a ``P\^ole
d'Attraction Interuniversitaire" (Belgium), by IISN-Belgium
(convention 4.4505.86)  and by the European Commission RTN programme
HPRN-CT-00131, in which G.B. is associated to K. U. Leuven.

\end{document}